\documentclass[aps,prl 
showpacs,
superscriptaddress,floatfix,reprint,citeautoscript,
twocolumn,
longbibliography,
]{revtex4-1}
\usepackage{graphicx}
\newcommand{\RNum}[1]{\uppercase\expandafter{\romannumeral #1\relax}}
\usepackage{amsmath}
\usepackage{amsmath}
\usepackage{amssymb}
\usepackage{bm}
\usepackage{float}
\usepackage{dcolumn}
\usepackage{subfigure}
\usepackage{units}
\usepackage{hyperref}
\usepackage[usenames]{color}
\usepackage{booktabs}
\usepackage{multirow}
\usepackage{siunitx}
\usepackage{wrapfig}
\usepackage{lipsum}
\usepackage{romannum} 
\usepackage{setspace}
\usepackage[dvipsnames]{xcolor}
\usepackage{xr}
\usepackage[normalem]{ulem}
\hypersetup{pdfborder=0 0 0,colorlinks=true,citecolor=blue,linkcolor=blue}
\input epsf

\begin{document}

\title{Magnetoelastic coupling enabled tunability of magnon spin current generation in 2D antiferromagnets} 
\author{N. Bazazzadeh}
\affiliation{Department of Physics, Shahid Beheshti University, Evin, Tehran 1983969411, Iran}

\author{M. Hamdi}
\altaffiliation[Present address:]{ Laboratory of Nanoscale Magnetic Materials and Magnonics (LMGN), Institute of Materials (IMX), School of Engineering (STI), EPFL, 1015 Lausanne, Switzerland.}
\email{mohamad.hamdi90@gmail.com}
\affiliation{Department of Physics, Shahid Beheshti University, Evin, Tehran 1983969411, Iran}
\newcommand{\MH}[1]{\textcolor{Green}{\uline{(Comment by MH: #1)}}}  

\author{S. Park}
\affiliation{Center for Correlated Electron Systems, Institute for Basic Science, Seoul 08826, Korea}
\affiliation{Department of Physics and Astronomy, Seoul National University, Seoul 08826, Korea}
\affiliation{Center for Theoretical Physics (CTP), Seoul National University, Seoul 08826, Korea}

\author{A. Khavasi}
\affiliation{Department of Electrical Engineering, Sharif University of Technology, Tehran, Iran}

\author{S. M. Mohseni}
\email{m-­mohseni@sbu.ac.ir}
\affiliation{Department of Physics, Shahid Beheshti University, Evin, Tehran 1983969411, Iran}

\author{A. Sadeghi}
\email{ali\_sadeghi@sbu.ac.ir}
\affiliation{Department of Physics, Shahid Beheshti University, Evin, Tehran 1983969411, Iran}
\newcommand{\AS}[1]{\textcolor{green}{Comment by AS: \sl#1}}  


\begin{abstract}
We theoretically investigate the magnetoelastic coupling (MEC) and its effect on magnon transport in two-dimensional antiferromagnets with a honeycomb lattice. MEC coefficient along with magnetic exchange parameters and spring constants are computed for monolayers of transition-metal trichalcogenides with N\'eel order ($\text{MnPS}_3$ and $\text{VPS}_3$)  and zigzag order ($\text{CrSiTe}_3$, $\text{NiPS}_3$ and $\text{NiPSe}_3$) by $ab$ $initio$ calculations. Using these parameters, we predict that the spin-Nernst coefficient is significantly enhanced due to magnetoelastic coupling. Our study shows that although Dzyaloshinskii-Moriya interaction can produce spin-Nernst effect in these materials, other mechanisms such as magnon-phonon coupling should be taken into account. We also demonstrate that the magnetic anisotropy is an important factor for control of magnon-phonon hybridization and enhancement of the Berry curvature and thus the spin-Nernst coefficient. Our results pave the way towards gate tunable spin current generation in 2D magnets by SNE via electric field modulation of MEC and anisotropy.


\end{abstract}

\maketitle

Recently, there is a growing interest in antiferromagnets (AFMs) as promising material platforms in spintronics. Because of the intrinsic timescale at THz range and the absence of the stray field, antiferromagnetic materials specially two dimensional (2D) AFMs \cite{jungwirth2016antiferromagnetic,jungfleisch2018perspectives,baltz2018antiferromagnetic} due to their reduced dimension are excellent candidates for high-speed and compact devices. 
Several interesting phenomena have been studied on AFMs such as the spin Hall effect \cite{ou2016strong,zhang2014spin,mendes2014large,kimata2019magnetic}, thermal Hall effect \cite{kawano2019thermal,mook2019thermal,kim2019magnon} and spin Nernst effect (SNE)\cite{cheng2016spin,zyuzin2016magnon,guo2017large,zhang2018spin,bazazzadeh2021symmetry}. In case of SNE, an experimental demonstration in $\text{MnPS}_3$ was first attributed to the presence of the Dzyaloshinskii-Moriya interaction (DMI) \cite{shiomi2017experimental}. However, the DMI in $\text{MnPS}_3$ was later found to be too small to be responsible for this effect \cite{wildes2021search}. In one hand, it is well known that the Berry curvature plays a crucial role on the transport of collective excitations in various systems \cite{nagaosa2010anomalous,xiao2010berry,chen2021anomalous}. On the other hand, a magnetoelastic wave (hybridized excitation of magnons and phonons~\cite{ogawa2015photodrive}), can carry large Berry curvature in the anticrossing regions between the magnon and phonon bands. These excitations which originate from the magnetoelastic coupling (MEC), can exhibit nontrivial topology even in the systems with trivial topology \cite{takahashi2016berry,zhang2019thermal,park2019topological}. Therefore, the MEC can be an important mechanism for inducing large SNE in AFMs \cite{park2020thermal}.

MEC has been considered in ferromagnetic materials with a square \cite{go2019topological} or honeycomb lattice \cite{thingstad2019chiral} and in AFMs with a square lattice \cite{zhang20203} in both magnon and phonon transport properties. Though these studies provide insight on the importance of the MEC in magnon transport, they do not deal with the realistic materials.
In this work, we choose transition-metal trichalcogenides (TMTCs) with honeycomb magnetic lattices because several magnetic phases especially AFM orders have been observed in these materials. 
Moreover, TMTCs are layered compounds with weak interlayer van der Waals interactions and are therefore excellent candidates for 2D AFMs~\cite{wiedenmann1981neutron,wildes1998spin,siberchicot1996band,sivadas2015magnetic,chittari2016electronic,joy1992magnetism,takano2004magnetic}.

We introduce a four-state method for calculating the MEC coefficients based on first-principles calculations in the framework of the density functional theory (DFT). 
Force constants are calculated by finite difference \cite{sarnthein1997origin} while the values for symmetric exchange interactions up to the third nearest neighbor
as well as the second nearest neighbor DMI interactions are taken from Ref.~\onlinecite{bazazzadeh2021symmetry}. Using these parameters we obtain the SNE coefficient for N\'eel ($\text{MnPS}_3$ and $\text{VPS}_3$) and zigzag ($\text{CrSiTe}_3$, $\text{NiPS}_3$ and $\text{NiPSe}_3$) ordered TMTCs. 
Our results show that including the MEC enhances the SNE coefficients by one or two orders of magnitude.
Furthermore, our study shows that magnetic anisotropy plays a crucial role in the magnon-phonon hybridization and thus SNE. These findings suggest a way for tunability and control of SNE generated spin current via modulation of MEC and anisotropy with e.g. applying a gate voltage.

\begin{figure}[t]
 \centering
 \includegraphics[width=0.45\textwidth]{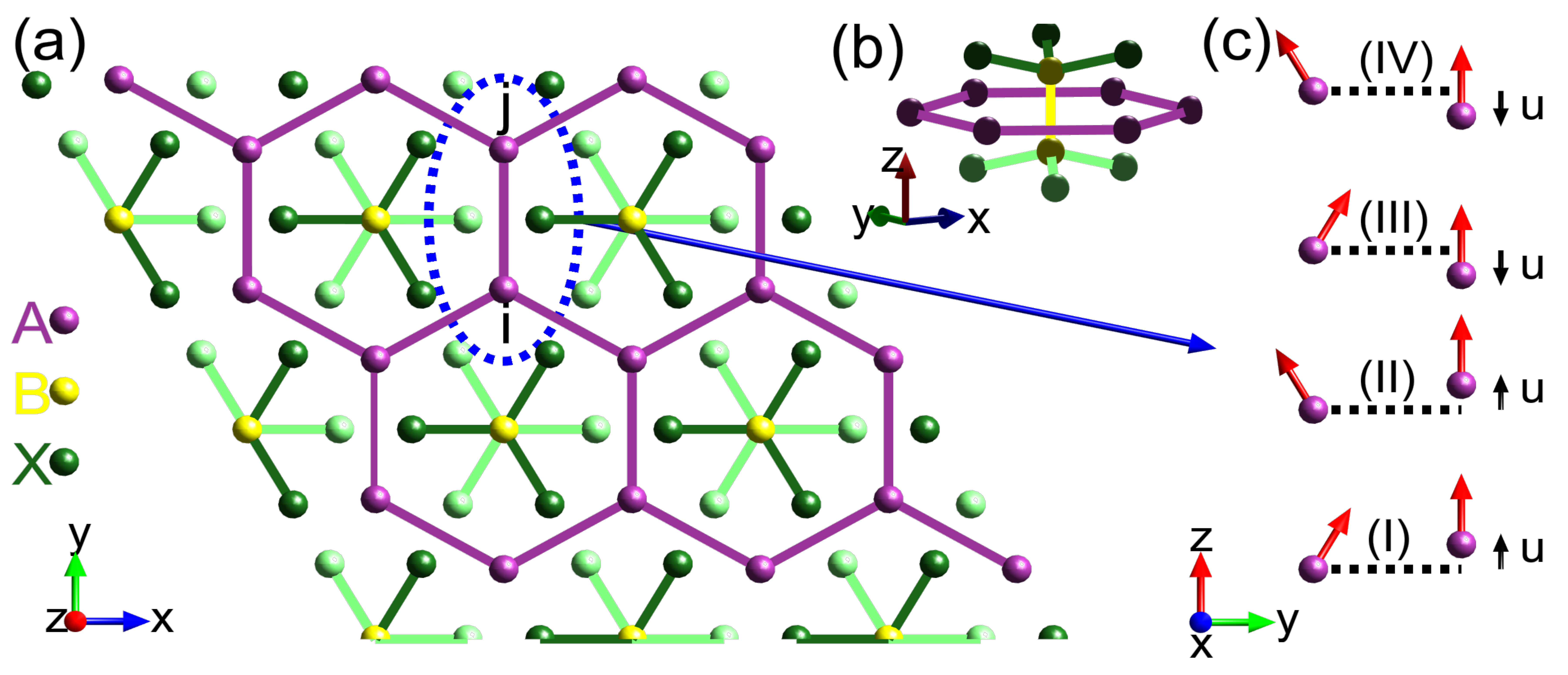}
 \caption{(a) Two-dimensional view of a $3\times3$ supercell representing a TMTC ($\text{ABX}_3$) material where A, B and X atoms are shown with purple, yellow and green spheres, respectively. (b) Side view of the material depicting $\text{B}_2\text{X}_6$ ligands located on the center of a hexagon made by A atoms. (c) Schematic representation  of the four states used to calculate the MEC coefficient. 
The big arrows represent the spin of two neighbouring magnetic atoms
while the smaller one indicates the out-of-plane displacement of one of them.}
 \label{TMTC}
\end{figure}

We perform DFT+U calculations as implemented in {\sc Quantum ESPRESSO}~\cite{giannozzi2009quantum}
with the Rappe-Rabe-Kaxiras-Joannopoulos ultrasoft (RRKJUS) pseudopotentials,
the Perdew-Burke-Ernzerhof (PBE) formalism of the generalized gradient approximation (GGA) \cite{perdew1996generalized}
and a kinetic energy cutoff of 400~eV. 
The Hubbard potential parameters are assumed as 5, 3.25, 4 and 6.45~eV for Mn, V, Cr and Ni, respectively~\cite{chittari2016electronic}. 
A tight convergence threshold of $10^{-7}$~eV is assumed for the total energy. 
A vacuum region larger than 20~\AA~  is used for each slab. 
Structural optimization are done while fixing the in-plane lattice constants to those reported in Refs.~\onlinecite{sivadas2015magnetic,chittari2016electronic}.
The magnetic ions A (Mn, V, Cr, Ni) make a honeycomb lattice with $\text{B}_2\text{X}_6$ ligands on the center of hexagons, as illustrated in Fig.~\ref{TMTC} (a-b).
We take a $3\times3$ supercell to make a separation between 
the periodic images of the deformed sites as large as $\sim 18$~\AA,
and a $k$-point grid of $3\times 3\times1$ for sampling the Brillouin zone.

In the investigated TMTCs with honeycomb magnetic lattice the spins possess N\'eel or zigzag ordering with out of plane spin orientation. The magnetic atoms can move out of plane and thus MEC comes into the play. Total Hamiltonian reads $H=H_\text{m}+H_\text{ph}+H_\text{mp}$. We write the magnetic Hamiltonian as 
\begin{equation} \label{eq:Hm}
\begin{split}
H_\text{m} & =J_1\sum\limits_{\langle {ij} \rangle}\mathbf{S}_i.\mathbf{S}_j+J_2\sum\limits_{\langle\langle {ij} \rangle\rangle}\mathbf{S}_i.\mathbf{S}_j
+J_3\sum\limits_{\langle\langle\langle {ij} \rangle\rangle\rangle}\mathbf{S}_i.\mathbf{S}_j\\&+D_2\sum\limits_{\langle\langle {ij} \rangle\rangle}\xi_{ij}\mathbf{\hat{z}}.( \mathbf{S}_i\times\mathbf{S}_j)+\mathcal{K}\sum\limits_{i}S^2_{iz}-\mu\sum\limits_{i}\mathbf S_{i}.\mathbf H\\
\end{split}
\end{equation}
where $J_{1}$, $J_{2}$ and $J_{3}$ are the Heisenberg exchange interaction parameters between the first, second, and third nearest neighbor spins, respectively. $D_2$ is the DMI parameter in the $z$ direction, $\mathbf{S}_i$ is the total spin at site $i$, and 
$\displaystyle \xi_{ij}=\text{sgn}\sum_{\langle{i,k}\rangle,\langle{k,j}\rangle}\mathbf{\hat{z}}
\cdot (\mathbf{r}_{ik}\times\mathbf{r}_{kj})$, where 
$\mathbf{r}_{ij}=\mathbf{r}_{j}-\mathbf{r}_{j}$ with $\mathbf{r}_{i}$ denoting the position of site $i$.
The last two terms in Eq.~(\ref{eq:Hm}) correspond to the easy-axis anisotropy 
and Zeeman coupling to external magnetic field, 
respectively.
We apply the Holstein-Primakoff transformation \cite{nolting2009quantum} for spin $\mathbf{S}_i=S^x_i\mathbf{\hat{x}}_i+S^y_i\mathbf{\hat{y}}_i+S^z_i\mathbf{\hat{z}}_i$
where $S^x_i=\sqrt{2S}(a_i+a^\dagger_i)/2$, $S^y_i=\sqrt{2S}(a_i-a^\dagger_i)/2i$ and $S^z_i=S-a_ia^\dagger_i$. 
The local spin coordinates for upward spins coincide the global coordinates
while for downward spins $\mathbf{\hat{x}}_i=\mathbf{\hat{x}}$, $\mathbf{\hat{y}}_i=-\mathbf{\hat{y}}$ and $\mathbf{\hat{z}}_i=-\mathbf{\hat{z}}$.

For each magnetic atom of mass $M$, only its out-of-plane displacements,  $u_i$, is relevant to the MEC~\cite{zhang20203}. The corresponding phonon Hamiltonian is therefore given by 
\begin{equation}  \label{eq:Hp}
H_\text{ph}=\dfrac{1}{2M}\sum\limits_{i} p_i^2+\dfrac{K_1^z}{2} \sum\limits_{\langle {ij}         \rangle}u_{ij}^2
\end{equation}
where $p_i$ denotes the linear momentum of atom $i$, $K_1^z$ is the nearest neighbor transverse spring constant and $u_{ij}=u_i-u_j$ is correspondingly the distance between the out-of-plane positions  
of  two nearest neighbors. Finally, the magnon-phonon coupling part of the Hamiltonian arising from MEC in an AFM honeycomb lattice is described to the linear order by~\cite{kittel1958interaction,go2019topological,thingstad2019chiral}
\begin{equation} \label{eq:Hmp}
H_\text{mp}=\kappa \sum_{\langle {ij} \rangle} \lambda_i (\mathbf S_i \cdot \hat{\mathbf r}_{ij})u_{ij}
\end{equation}
where $\kappa$ is the magnetoelastic constant and $\hat{\mathbf r}_{ij} = {\mathbf r}_{ij}/r_{ij}$. Note that Eq.~(\ref{eq:Hmp}) is valid only if $\lambda_i= S_i^z/S \approx \pm 1$.

The numerical values of coefficients in Eqs.~(\ref{eq:Hm}), (\ref{eq:Hp}) and (\ref{eq:Hmp}) 
are listed in Table~\ref{table}. The values of $J_1$, $J_2$, $J_3$ and $D_2$ are those previously determined from first principles calculations by a four-state method Ref.~\onlinecite{bazazzadeh2021symmetry}. $K_1^z$ is calculated by finite difference method from DFT+U energies. For calculating $\kappa$, we adopt the following four-state method.
In short, all the spins in the supercell are set parallel to the $z$-axis while the spin of atom at a given site $i$ is slightly tilted toward or away from a nearest neighbor atom at site $j$ which is displaced normal to the sheet plane either upward or downward by a tiny value, 
e.g. $u=0.05$~\AA.
 The four possible configurations of a pair of atoms highlighted in Fig.~\ref{TMTC}(a) are:
\begin{table*}[!t]
 \centering
 \caption{Magnetic ground state, lattice constant (\AA), 
magnetic moment per TM atom,   
 exchange (meV) and DMI parameters ($\mu$eV), 
  N\'eel temperature (K) from Monte Carlo simulations and experiment,
 anisotropy coefficient (meV), transverse spring constant (meV/\AA$^2$) and MEC coefficient (meV/\AA) 
 for  the five investigated TMTCs.
  The two columns on the right are calculated in this work.
  The first eight columns are adopted from Ref.~\cite{bazazzadeh2021symmetry}.
  $\mathcal{K}$ (meV) is set to the available experimental values for bulk
  $\text{MnPS}_3$~\cite{wildes1998spin} and $\text{NiPS}_3$~\cite{lanccon2018magnetic}.
For $\text{NiPSe}_3$ and $\text{VPS}_3$ no experimental  $\mathcal{K}$ is available 
while the experimental values for bulk $\text{CrSiTe}_3$~\cite{williams2020magnetic} makes our 2D calculation numerically unstable
    (We note that the GS 2D $\text{CrSiTe}_3$ is zigzag~\citep{sivadas2015magnetic} while it is ferromagnetic in bulk~\cite{williams2020magnetic}).
    Therefore, for $\text{NiPSe}_3$ we use the same  $\mathcal{K}$ value as $\text{NiPS}_3$, 
    and for $\text{VPS}_3$ and $\text{CrSiTe}_3$ we took the smallest values that lead to stable numerical solutions.}
 \begin{tabular*}{1\linewidth}{@{\extracolsep{\fill}}lcccccccccccc}
 \hline \hline
  \multicolumn {4}{c}{Structure} & \multicolumn{4}{c}{Exchange and DMI} & \multicolumn{2}{c}{N\'eel Temp.} 
   
\\
\cline{1-4} \cline{5-8} \cline{9-10}  

 Material&GS & $a$ & $S / \mu_B$&  $J_1$ &$J_2$&$J_3$&$|D_{2z}|$ &MC &Exp &$\mathcal{K}$&  $K^z_1$&$|\kappa|$ \\
 \cline{1-1}\cline{2-4} \cline{5-7} \cline{8-8} \cline{9-10}\cline{11-11} \cline{12-12}\cline{13-13}
 $\text{MnPS}_3$   &N\'eel&5.88&4.56&0.527 &0.024 &0.150&0.39 &150&    78\cite{wildes1998spin}  &-0.002&479.9&0.0292\\  
 $\text{VPS}_3$    &N\'eel&5.85&2.82&7.387 &0.068 &0.223&7.23 &530&  -                          &-0.014&106.7&0.6858\\
 $\text{CrSiTe}_3$ &zigzag&6.84&3.76&-0.990&0.009 &0.389&39.00&48 & 80\cite{lin2016ultrathin}   &-0.023&118.9&0.9320\\
 $\text{NiPS}_3$   &zigzag&5.82&1.58&-1.039&-0.163&3.882&4.54 &105& 150\cite{kim2019suppression}&-0.190&243.6&0.2949\\ 
 $\text{NiPSe}_3$  &zigzag&6.14&1.56&-1.131&-0.069&3.975&43.90&110&  -                          &-0.190&173.8&1.9729\\
 \hline
 \end{tabular*}

 \label{table}
 
\end{table*}

\begin{equation}\label{4state}
    \begin{cases}
\begin{array}{lll}
 \text{state I:  } & \mathbf{S}_i=(0,+S_y,S_z)  &  \text{  and      $u_j=+u$} \\
 \text{state II: } & \mathbf{S}_i=(0,-S_y,S_z)  &  \text{  and      $u_j=+u$} \\
 \text{state III:} & \mathbf{S}_i=(0,+S_y,S_z)  &  \text{  and      $u_j=-u$} \\
 \text{state IV: } & \mathbf{S}_i=(0,-S_y,S_z)  &  \text{  and      $u_j=-u$} 
\end{array}
    \end{cases}
\end{equation}
as illustrated schematically in Fig.~\ref{TMTC}(c). A value of $S_y = \pm 0.3S$ which leads to $\lambda_i=\pm 0.954$, fulfills the validity requirements for Eq.~(\ref{eq:Hmp}) as discussed before. It can be easily shown that the MEC coefficient reads 
\begin{equation} \label{MEC}
\kappa=\frac 1 {4S_y u}\left( E_{\text{II}}+E_{\text{III}}-E_{\text{I}}-E_{\text{IV}} \right).
\end{equation}
where  $E_\text{I}$, $E_\text{II}$, $E_\text{III}$ and $E_\text{IV}$ are the DFT+U energies of the four states defined in Eq.~(\ref{4state}). 

To obtain the band structure of the excitations the total  bosonic BdG  Hamiltonian in the Fourier space, 
\begin{figure*}[!t]
\centering
\includegraphics[width=0.99\textwidth]{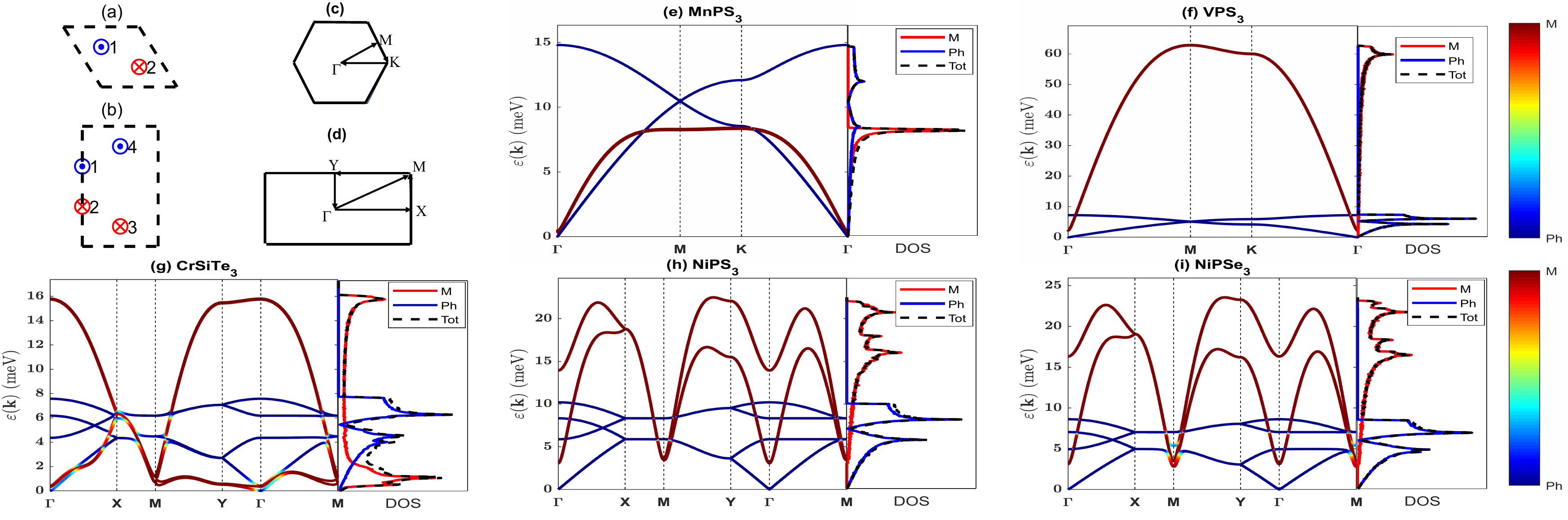}
\caption{Magnetic unit cell for (a) the N\'eel phase and (b) zigzag phase. Symmetry points for (c) the N\'eel phase and (d) zigzag phase. The band
structure and density of states (DOS) of (e) $\text{MnPS}_3$, (f) $\text{VPS}_3$, (g) $\text{CrSiTe}_3$, (h) $\text{NiPS}_3$ and (i) $\text{NiPSe}_3$. The spin character $|\langle S^z \rangle|$ of the modes is indicated with colors.}
\label{magnonband}
\end{figure*}  
\begin{equation}
H = \frac{1}{2}\sum_\mathbf k \phi^\dagger_\mathbf k\mathcal{H}(\mathbf k)\phi_\mathbf k
\end{equation}
is diagonalized using the Colpa's method~\cite{bazazzadeh2021symmetry,colpa1978diagonalization,park2019topological} (for details, see SM~\cite{sup})
\begin{equation} 
\Psi^\dagger(\mathbf k) \mathcal{H}(\mathbf k) \Psi(\mathbf k)=E(\mathbf k)
\end{equation}
The elements of $2N \times 2N$ diagonal matrix $E(\mathbf k)$ are 
the eigenenergies while the associated eigenstates are the columns of $\Psi(\mathbf k)$ and satisfy $\hat \varphi \mathcal{H}(\mathbf k) |n,\mathbf k\rangle =E_{nn}(\mathbf k) |n,\mathbf k\rangle$ and $\left \langle n,\mathbf k|\hat \varphi |m,\mathbf k\right \rangle=\hat \varphi_{nm}$.
Note that $\Psi(\mathbf k)$ is a paraunitary matrix i.e. it satisfies $\Psi^\dagger(\mathbf k) \hat{\varphi}=\hat{\varphi} \Psi^{-1}(\mathbf k) $,
 where $\hat{\varphi} = \text{diag} (1,\cdots,1,-1,\cdots,-1) $.
For the N\'eel and zigzag orders, shown in Figs.~\ref{magnonband}(a-b),
 $N=4$ and 8, respectively.

Since the spin is not conserved in the presence of MEC, the semiclassical approach is not valid and linear response theory should be applied \cite{park2020thermal}. 
For this purpose, a generalized Berry curvature should be defined for operator $\mathbf{\theta}=\frac{1}{4}(S^z\hat{\varphi}\mathbf v+\mathbf v \hat{\varphi}S^z)$ where $\mathbf v=\hbar^{-1}\partial_\mathbf{k} H$ 
is the velocity operator and $S^z=-\sum_i^{N/2} \lambda_i a^\dagger_i a_i$ is the magnon spin operator.
The generalized Berry curvature of the $n$th band is given by~\cite{li2020intrinsic} 
\begin{equation}
(\Omega_n^{\theta}(\mathbf k))_{xy}= \sum_{m\neq n}
\frac{2i\left \langle n|\theta_x |m\right \rangle \left \langle m|v_y |n\right \rangle (\hat{\varphi})_{nn} (\hat{\varphi})_{mm}}{((\hat{\varphi}E)_{nn}-(\hat{\varphi}E)_{mm})^2}\\ 
\label{berry}
\end{equation}
where the index $\mathbf k$ is dropped for simplicity (the conventional Berry curvature is presented in the SM~\cite{sup}.) 
The SNE coefficient is defined as $\alpha^s_{xy}\equiv \mathbf j_{SNE}/(\hat z\times \nabla T)$ \cite{cheng2016spin}, where $\nabla T$  is an in-plane temperature gradient and the magnon spin current, $\mathbf j_{SNE}$ is calcauted based on the linear response theory~\cite{park2020thermal,li2020intrinsic} as 
\begin{equation}
\label{jn}
\alpha^s_{xy} = 2k_B\sum_{n=1}^N\int d\mathbf{k}(\Omega_n^{\theta}(\mathbf k))_{xy} c_1\left(\frac{E_{nn}}{k_B T}\right)
\end{equation}
where $c_1(x)=[1+\rho(x)]\log[1+\rho(x)]-\rho(x)\log \rho(x)$ and $\rho(x) =1/(e^x-1)$. Note that only the particle bands (the first $N$ bands) contribute to the summation.

In an experimental report a bulk crystal of $\text{MnPS}_3$ was used whose edges are cut at an inclination angle  $\eta=45^\circ$
and Pt stripes were used as electrodes~\cite{shiomi2017experimental}. 
To get results comparable with such SNE measurement in TMTCs,
we adopt the SNE signal introduced in Ref.~\onlinecite{bazazzadeh2021symmetry} 
\begin{equation}
\mathcal{S}_{SNE}=\dfrac{V_{SNE}}{\Delta T}=\frac{e}{\hbar d}\, \rho\theta_{SH} \,\alpha^s_{xy}\cos\eta.
\label{SNESignal}
\end{equation}
where $V_{SNE}$ is voltage due to inverse spin Hall effect, 
$\Delta T$ is the temperature difference across the length of Pt stripes, 
$d$ is the inter-layer distance in the TMTC,
and $\rho=10^{-6}~{\Omega}$/m and $\theta_{SH}=0.15$ are the electrical resistivity and the spin Hall angle of Pt, respectively.

 All the parameters for constructing the Hamiltonian,
 including the calculated MEC coefficients $\kappa$ and  transverse spring constant $K^z_1$,
 are reported in Table ~\ref{table}.
 The resulting band structure along the high symmetry directions shown in Figs.~\ref{magnonband}(c) and (d) as well as the density of states (DOS) of the quasiparticle bands
 are presented in Figs.~\ref{magnonband}(e-f) and (g-i) for the N\'eel and zigzag ordered materials, respectively.
 From the absolute value of the z-component of spin, $|\langle S^z \rangle|$ ~\cite{sup}, encoded by the color bar in the band structure plots, one clearly identifies two degenerate magnon bands (red) and two phonon bands (blue) in the N\'eel phase. On the other hand, four magnon bands (two doubly-degenerate bands) and four phonon bands are identified for the zigzag phase. 

\begin{figure*}
	\centering
	\includegraphics[width=0.99\textwidth]{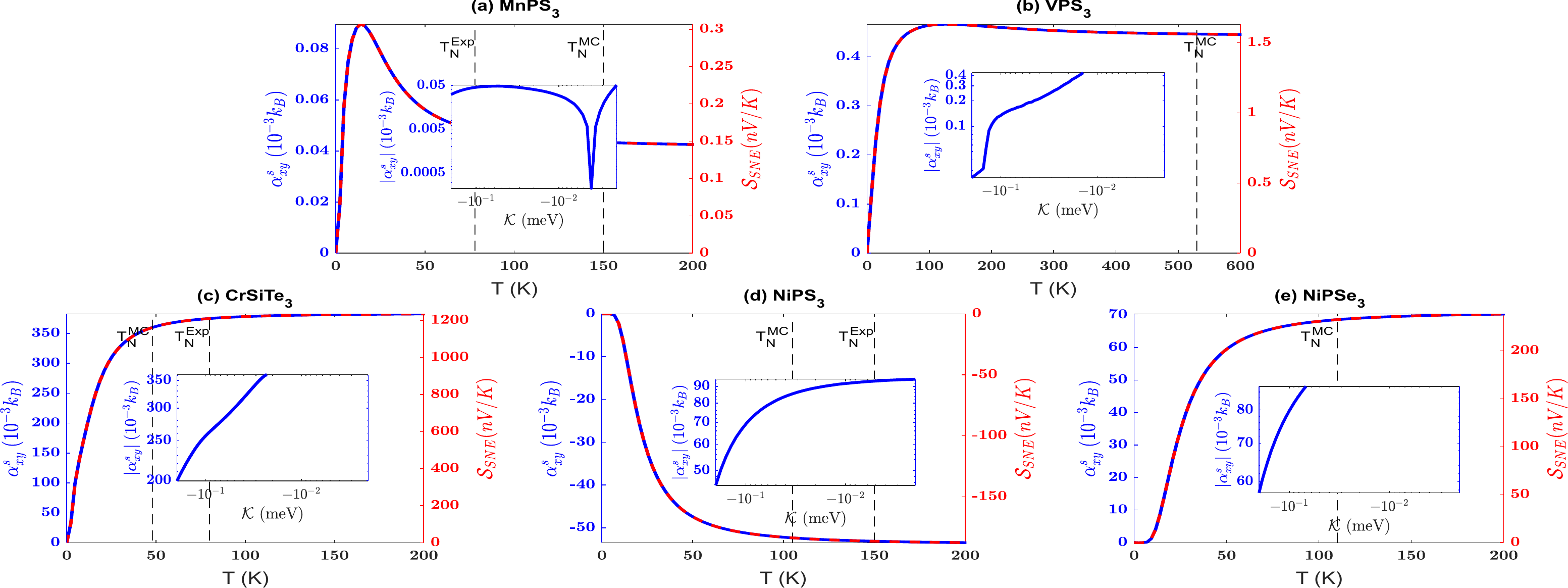}
	\caption{The SNE coefficient (blue dashed lines) and SNE signal (red dashed lines) versus temperature, top for N\'eel phase and bottom for zigzag phase. The experimental and Monte Carlo N\'eel temperatures are also marked with vertical black dashed lines. Insets: The SNE coefficient in terms of easy-axis anisotropy $\mathcal{K}$ plotted in logrithmic scale.}
	\label{SNE}
\end{figure*}

As can be seen in Fig. \ref{magnonband},
MEC leads to hybridization of magnon and phonon bands which manifests itself as the anticrossing regions (some of anticrossing regions are shown in Fig. S6). Therefore, the generalized Berry curvature  becomes very large in these anticrossing regions for both N\'eel and zigzag ordered materials, as illustrated in Figs. S3 and S4 of the SM~\cite{sup}. Consequently, the SNE coefficient is largely increased as seen in Fig.~\ref{SNE}.
By including the MEC in the calculations, the maximum SNE coefficient within the shown temperature range is enhanced by a factor of $\sim$23, 12, 13, 620 and 99 for $\text{MnPS}_3$, $\text{VPS}_3$, $\text{CrSiTe}_3$, $\text{NiPS}_3$ and $\text{NiPSe}_3$, respectively.

A log-log plot of the absolute value of SNE coefficient, $|\alpha^s_{xy}|$, as a function of the anisotropy $\mathcal{K}$ at $T=50$ K is also depicted in the insets of Fig.~\ref{SNE} (Note that there is a sign change at $\mathcal{K}=-0.004$~meV for $\text{MnPS}_3$). It is clearly seen that the SNE coefficient is fairly sensitive to changes in the anisotropy. This is due to the effect of anisotrpy in displacing bands and thus anticrossing regions which in turn changes the generalized Berry curvature and SNE.
 
There are also other factors that affect SNE coefficient. For $\text{MnPS}_3$ and $\text{VPS}_3$, there are anticrossing region at higher energies (around 8~meV) leading to different behavior at higher temperatures (around 90~K). 
The situation is more complicated for zigzag ordered materials due to magnon-phonon hybridization in several regions. 
Specially, for $\text{CrSiTe}_3$ we can see large generalized Berry curvature almost everywhere in the Brillouin zone (Fig. S4)
which leads to very large SNE coefficient compared to other materials. 
For $\text{NiPS}_3$ and $\text{NiPSe}_3$, we see anticrossing regions around the M and $\Gamma$ points (See Figs. \ref{magnonband}(h) and (i)). However, relatively small generalized Berry curvature is induced near the $\Gamma$ point (see Fig. S4). We thus consider only the anticrossing regions around the M point. For both materials the  hybridization is stronger at the anticrossing with lower energies.
However, since the MEC coefficient $\kappa$ of $\text{NiPSe}_3$ is very large, the magnon bands which carry $+1$ and $-1$ spins and the phonon bands strongly mix at this region (around the M point). This results in magnon bands with almost no spin, i.e. the spin sectors with opposite spins cancel each other (shown in Figs. S5 (n) and (o) where the lower bands around the M point have considerable magnon contents but almost no spin), so they have less contribution to the SNE coefficient. 
On the other hand,  $\kappa$ of $\text{NiPS}_3$ is much smaller, thus there are still bands that can carry spin at the anticrossings around the M point with lower energies. 
So the lower bands mainly contribute to the SNE coefficient (see Fig. S4), and their contribution is negative leading to negative SNE coefficient.
In contrast, the anticrossings around the M point with higher energies (higher bands at it can be seen in Fig. S4) contributes mainly to the SNE coefficient of $\text{NiPS}_3$ and their contribution is positive. 

We also note that the individual contribution of each band to SNE coefficient can be orders of magnitude larger than the final SNE (see Fig. S7). However, the sum of positive bands is almost equal to negative bands but opposite in sign (see the inset of Fig. S7). So there is very small difference between the absolute values of positive and negative bands which leads to very small SNE coefficient compared to the contribution of each band yet much larger than SNE coefficient in the absence of MEC. In contrast, when $\kappa = 0$ the contribution of each band is comparable with final SNE coefficient.

In summary, this study reveals that MEC induces large generalized Berry curvatures in anticrossing regions which in turn translates into large SNE in TMTC materials. We also found that the SNE is sensitive to anisotropy and paves the way for designing tunable spintronic devices. Among the TMTC materials studied in this work, $\text{NiPSe}_3$ has the largest MEC, but the largest SNE is observed in $\text{CrSiTe}_3$. Despite the focus of previous studies, presence of DMI cannot explain large SNE observed in $\text{MnPS}_3$ \cite{shiomi2017experimental}. Other magnon-phonon coupling mechanisms should be also considered in the future works to better understand the topological properties of TMTC antiferromagnets. 

\textbf{Acknowledgment}. This project is funded by Iran Science Elites Federation (ISEF). The support and computatioanl resources from the Center for High Performance Computing (SARMAD) at SBU is gratefully acknowledged. S. P. was supported by the Institute for Basic Science in Korea (Grant No. IBS-R009-D1), 
Samsung  Science and Technology Foundation under Project Number SSTF-BA2002-06,
the National Research Foundation of Korea (NRF) grant funded by the Korea government (MSIT) (No.2021R1A2C4002773, and No. NRF-2021R1A5A1032996).

\bibliography{ref}
\end{document}